\documentclass[prl,twocolumn,aps,superscriptaddress,showpacs]{revtex4}

\usepackage{amsfonts}
\usepackage{mathrsfs}
\usepackage{amsmath}
\usepackage{color}
\usepackage{graphicx}
\usepackage{bm}
\usepackage{amssymb}
\usepackage{xspace}
\usepackage{epstopdf}
\usepackage{dcolumn}
\usepackage{longtable}
\usepackage{multirow}
\usepackage[colorlinks=true, letterpaper=true, pdfstartview=FitV, linkcolor=blue, citecolor=blue, urlcolor=blue]{hyperref}

\begin{document}

\title{Graphdiyne Nanoribbons with Open Hexagonal Rings: Existence of Topological Unprotected Edge States}

\author{Cong Chen}
\affiliation{Department of Physics, Key Laboratory of Micro-nano Measurement-Manipulation and Physics (Ministry of Education), Beihang University, Beijing 100191, China}

\author{Jin Li}
\affiliation{Laboratory for Quantum Engineering and Micro-Nano Energy Technology and School of Physics and Optoelectronics, Xiangtan University, Hunan 411105, China}


\author{Xian-Lei Sheng}
\email{xlsheng@buaa.edu.cn}
\affiliation{Department of Physics, Key Laboratory of Micro-nano Measurement-Manipulation and Physics (Ministry of Education), Beihang University, Beijing 100191, China}


\begin{abstract}
Using first principles calculations, we  studied a new class of graphdiyne nanoribbons (GDYNR)  with open hexagonal rings on the edges.
 To avoid the effects from dangling bond, hydrogen or oxygen atoms were absorbed
on the edges. There are two kinds of GDYNR depending on the edge structures, armchair and zigzag. The electronic structures show that all of them are semiconductors. The band gap can be tuned by the width of GDYNR. As the  width of nanoribbons increases, the energy gap decreases firstly and then increases, and  reaches a minimum gap for both  kinds. To understand the intriguing phenomenon, we constructed a tight-binding model for GDYNR and found that the existence of the minimum of the energy gap is due to the competition between the interaction within the two edges and the coupling in between.  Furthermore, topological unprotected edge states are found in the band structure of a semi-infinite system by calculating surface Green's function.
If GDYNR could be synthesized in experiments, it would be useful for the nanodevices in the future.
\end{abstract}

\maketitle


\section{introduction}
\label{introduction}
The electronic properties of novel carbon materials have been studied extensively including 0-dimensional (0-D) fullerene~\cite{C601985,Schultz1965,PhysRevB.78.201401,ShengXL_PCCP2009}, 1-D nanotube~\cite{Iijima1991nature,saito1998,SHENG201063}, 2-D graphene and graphyne~\cite{RevModPhys.81.109,PhysRevB.75.041401,Octagraphene,Graphdiyne,ZhouJ_JCP2011,QiuHH_PLA,QiuHH_PhysicaB}, and 3-D diamond and other carbon allotropes ~\cite{ANIE:ANIE201600655, ANIE:ANIE201400131,PhysRevLett.102.175506,T-carbon,CuiHJ_BCTC8,CuiHJ_CMS2015}, due to the strong ability of carbon element to form countless network structures with $sp$-, $sp^2$- and $sp^3$-hybridized chemical bonds. Graphene and diamond are typical carbon allotropes with $sp^2$- and $sp^3$-hybridization, respectively. Graphyne (graphdiyne) is a kind of $sp$- and $sp^2$-hybridized carbon materials, by inserting one acetylenic (diacetylenic) linkages between nearest hexagonal rings in graphene lattice~\cite{PhysRevB.58.11009}. Recently, graphdiyne films have been successfully synthesized via cross-linking reaction using hexaethynylbenzene on top of copper surface\cite{Graphdiyne}, which shows typical semiconducting character by both of experiments~\cite{Graphdiyne} and theoretical predictions~\cite{PhysRevB.58.11009,PanLD2011}.
Since graphdiyne is a two-dimentional carbon sheet as well as graphene, it attracts extensive attentions on its electronic, transport, and mechanical properties. For example, graphdiyne was reported in experiment as a metal-free material as hole transfer layer to fabricate quantum dot-sensitized photocathodes for hydrogen production~\cite{jacs.5b12758}. It was also predicted as a promising material for detecting amino acids~\cite{Chen2015}. Moreover, there are also other kinds of graphyne with direction dependent Dirac cones~\cite{PhysRevLett.108.086804}.

It is well known that graphene nanoribbons (GNR) hold rich electronic, magnetic and transport properties, and the band gap can be tuned by cutting GNR with different width and edges (zigzag or armchair) for the configuration~\cite{RevModPhys.81.109}. Thus, it is natural to ask how the properties of  graphdiyne nanoribbons (GDYNR) behave. GDYNR could have much more configurations than GNR, because of the complexity of graphdiyne network. Based on their different edge structures, GDYNR could be benzene ring or acetylene terminated edges, with controllable band gaps from 0.5 to 1.3 eV \cite{PanLD2011,ShuaiZG_ACSnano,C1RA00481F}, larger than the band gap of graphdiyne sheet of  about 0.5 eV. Is it possible to cut a kind of nanoribbons with smaller band gap than bulk, and is there any edge state?


We will answer this question in this work by cutting a different kind of GDYNR. Based on the projected density of states (PDOS) analysis, we find that hexagonal rings in graphdiyne play an important role in the electronic states around Fermi level. Therefore, it would be interesting to investigate the electronic properties of GDYNR with open hexagonal rings on the edges.
Here, we designed a family of such GDYNR, which is quite different from the  closed carbon hexagons edged GDYNR \cite{PanLD2011,ShuaiZG_ACSnano,C1RA00481F}.
The former can be semiconductor or semimetal while the later can only be semiconductor. The results show that both of armchair and zigzag GDYNR can be a semimetal with nearly zero band gap or semiconductor with band gaps  around 0.5 eV.
 The valence and conductive states are mainly localized on the edges, and Dirac-like dispersion appears in a small energy window around the Fermi energy. Furthermore, the band gap can be tuned by varying the width of GDYNR. When the  width of nanoribbons increases, the energy gap decreases firstly and then increases, and  reaches a minimum gap  for both of the armchair and zigzag GDYNR. To understand the intriguing phenomenon, we constructed a tight-binding model for GDYNR and found that the interaction between the edges plays a key role. When the edge interaction is zero, the model corresponds to a GDYNR with infinite width, whose energy gap is only depending on the edge properties. As the interaction increases, it corresponds to decrease the GDYNR width, the energy gap decreases firstly and then increases, in good agreement with the first-principles results. To see the electronic properties of a single edge, we calculate the band dipersion of an semi-infinite system using surface Green's method, and find topological unprotected edge states.

\section{Computational Method}
Most of the calculations were performed within density-functional theory (DFT) as implemented within  Vienna \textit{ab initio} simulation package (VASP) \cite{VASP1,VASP2} with the projector augmented wave (PAW) method \cite{paw}. The generalized gradient approximation (GGA) with Perdew-Burke-Ernzerholf (PBE) \cite{GGAPBE} formalism was adopted for the exchange correlation potential. The plane-wave cutoff energy was taken as 400 eV. The supercells were used to simulate the isolated nanoribbons and the distance of nanoribbons is larger than 12 {\AA} in order to avoid interactions. The Monkhorst-Pack scheme was used to sample the Brillouin zone~\cite{PhysRevB.13.5188}, and a mesh of 1$\times$1$\times$15 k-point sampling was used for the calculations. The geometries were optimized when the remanent Hellmann-Feynman forces on the ions are less than 0.01 eV/{\AA}. To further confirm the calculations and get more data for wide GDYNR, we performed first-principles calculations by using the software package OpenMX~\cite{openmx}, and the results of the two softwares are in good agreement with each other.

\section{Results and disscusion}

\begin{figure}[tbp]
\includegraphics[width=1.0\linewidth,clip]{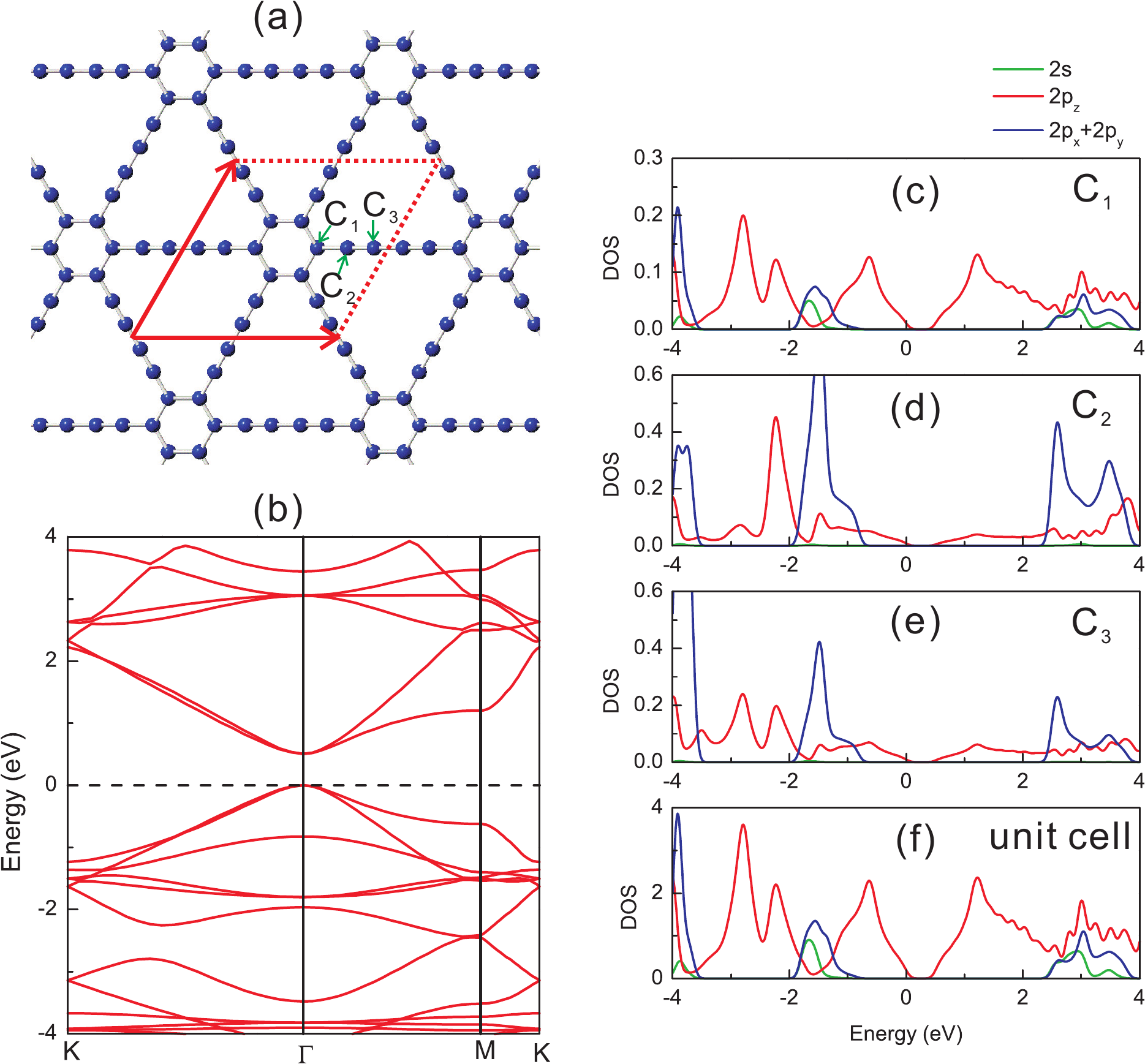}
\caption{(a) Schematic structures of graphdiyne, the unit cell (red rhombic box) and the three non-equivalent atoms $C_1$, $C_2$ and $C_3$. (b) Electronic energy bands of graphdiyne. (c), (d) and (e) is the projected density of states (PDOS)of atoms $C_1$, $C_2$ and $C_3$, respectively. (f) is the total PDOS of a unit cell. The PDOS of each orbital is shown as different color line.}
\label{fig:sheet}
\end{figure}

The graphdiyne sheet was optimised firstly and the relaxed lattice constant is  9.46 {\AA}, in good agreement with the previous theoretical values of 9.44 {\AA}\cite{PhysRevB.58.11009} and 9.48 {\AA}\cite{ShuaiZG_ACSnano}. The point group symmetry of graphdiyne is $C_{6v}$. There are three non-equivalent atoms with Wyckoff coordinates (0.1516, 0, 0), (0.2994, 0, 0), and (0.4292, 0, 0). Before studying the electronic properties of GDYNR, we first calculated the electronic structure of the two-dimensional graphdiyne as shown in Fig.~\ref{fig:sheet}. Fig.~\ref{fig:sheet} (a) shows the graphydiyne sheet structure with a unit cell in the red rhombic box and three non-equivalent atoms are labeled as C$_1$, C$_2$ and C$_3$, respectively. Fig.~\ref{fig:sheet} (b) is the electronic band structure of graphdiyne. To understand the contributions of each atom, PDOS was calculated. Figs.~\ref{fig:sheet} (c-f) are the PDOS of atom $C_1$, $C_2$, $C_3$ and the total DOS  of a unit cell. All of the results indicate that $2p_z$ orbital is the most important around Fermi energy, quite similar to graphene. Thus, a $\pi$-electron model can ben constructed to describle the low energy physics~\cite{Cui2013}. To further check the band structure by different computational methods, we also calculated the band structure of graphdiyne by HSE hybrid exchang correlation potential, giving a direct energy gap of 1.1 eV and quite similar band dispersion. Hereafter, we only calculated the electronic states of nanoribbons at PBE level.  

%
%
\begin{figure}[tbp]
\includegraphics[width=0.8\linewidth,clip]{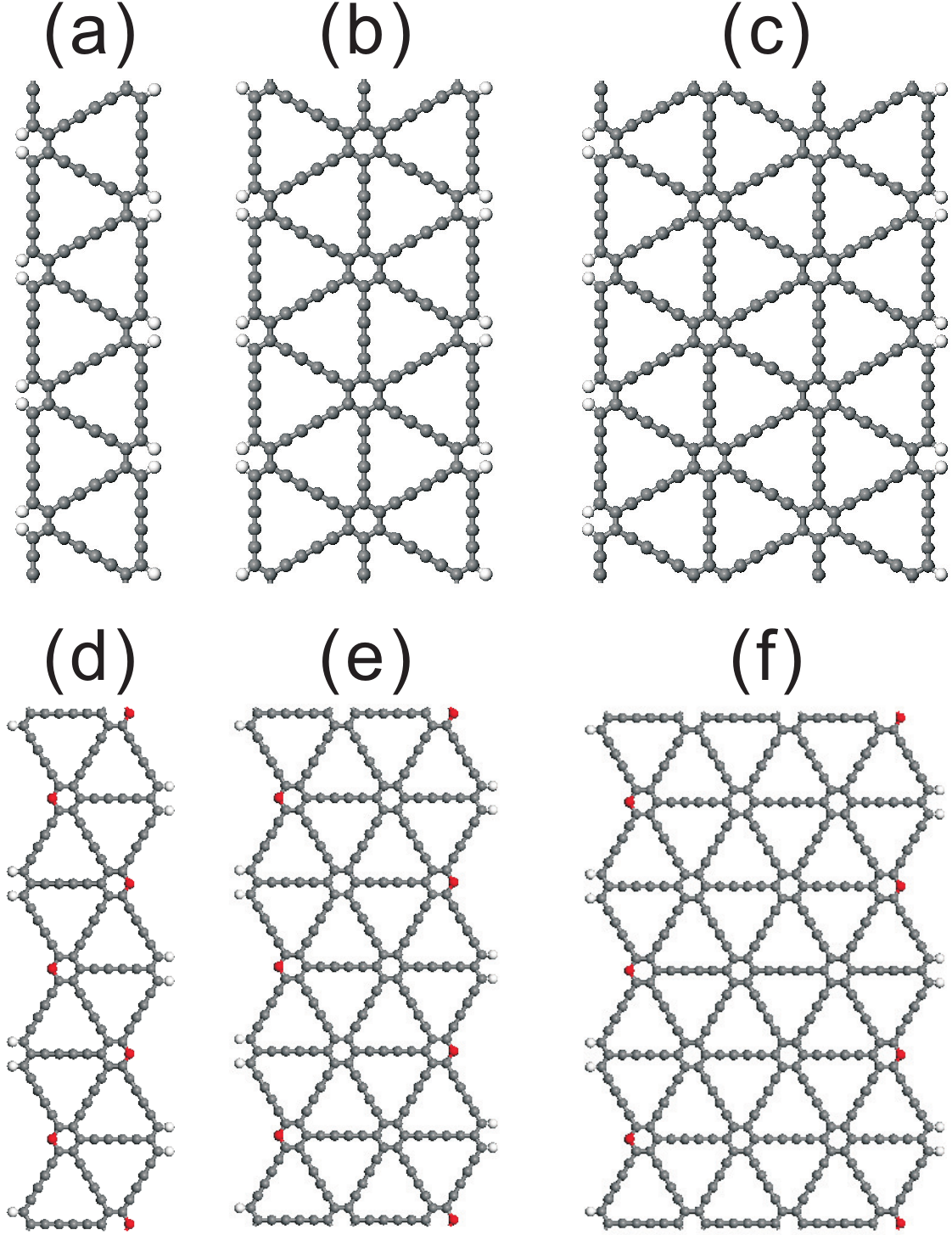}
\caption{ Schematic structures of the two kinds of graphdiyne nanoribbons (GDYNR): upper panels for armchair and lower panels for zigzag. To avoid dangling bonds, hydorgen (white ball) or oxygen (red ball) atoms are absorbed. }
\label{fig:structure}
\end{figure}

The structure of graphdiyne is quite different from graphene for the existence of $sp$ hybridized acetylenic linkages. Therefore, the nanoribbon structure is more complicated than GNR. For GNR, there are only two kinds of structures, armchair and zigzag. Graphdiyne consists of carbon hexagons and atom lines, so that the edges of nanoribbons could be closed hexagons, open hexagons or dangling atom lines. The first class has been studied already with semiconducting character~\cite{PanLD2011,ShuaiZG_ACSnano,C1RA00481F}. The third calss ofen have mixed electronic states of  dangling states, difficult to get a clear conclusion.  Therefore,  we focus on the nanoribbons with open hexagonal edges as shown in Fig.\ref{fig:structure}. We cut graphdiyne along two perpendicular directions, respectively, to get armchair nanoribbons [see Fig.\ref{fig:structure} (a-c)] and zigzag nanoribbons [see Fig.\ref{fig:structure} (d-f)], respectively. For convenience, we label GDYNR as armchair/zigzag GDYNR-$n$ ($n=2,3,4,...$) for armchair/zigzag nanoribbon which has $n$ atom lines along armchair/zigzag direction. There is no GDYNR-$1$, because in this case the nanoribbon would be disconnected when the hexagons do not close. Fig.\ref{fig:structure} shows the first three structures of armchair GDYNR and zigzag GDYNR, respectively. In this work, we calculated both GDYNR from $n=2$, and upto $n=15$ which is large enough to avoid edge interaction.

%

\begin{figure}[tbp]
\includegraphics[width=1.0\linewidth,clip]{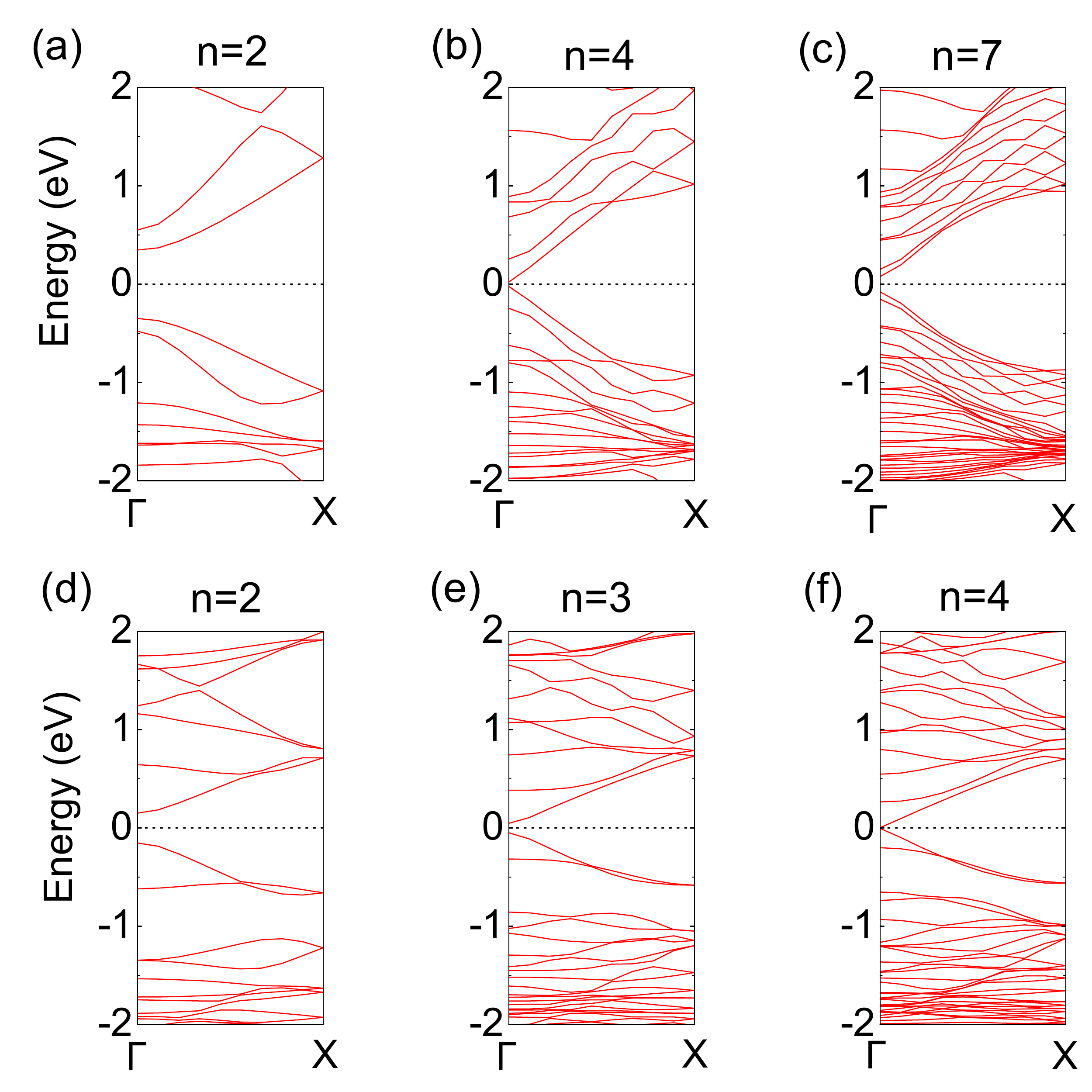}
\caption{(a)-(c) Band structures of armchair GDYNR with $n=2,4,7$, respectively. (d)-(f) Band structures of zigzag GDYNR with $n=2,3,4$, respectively. }
\label{fig:band}
\end{figure}
\begin{figure}[tbp]
\includegraphics[width=1.0\linewidth,clip]{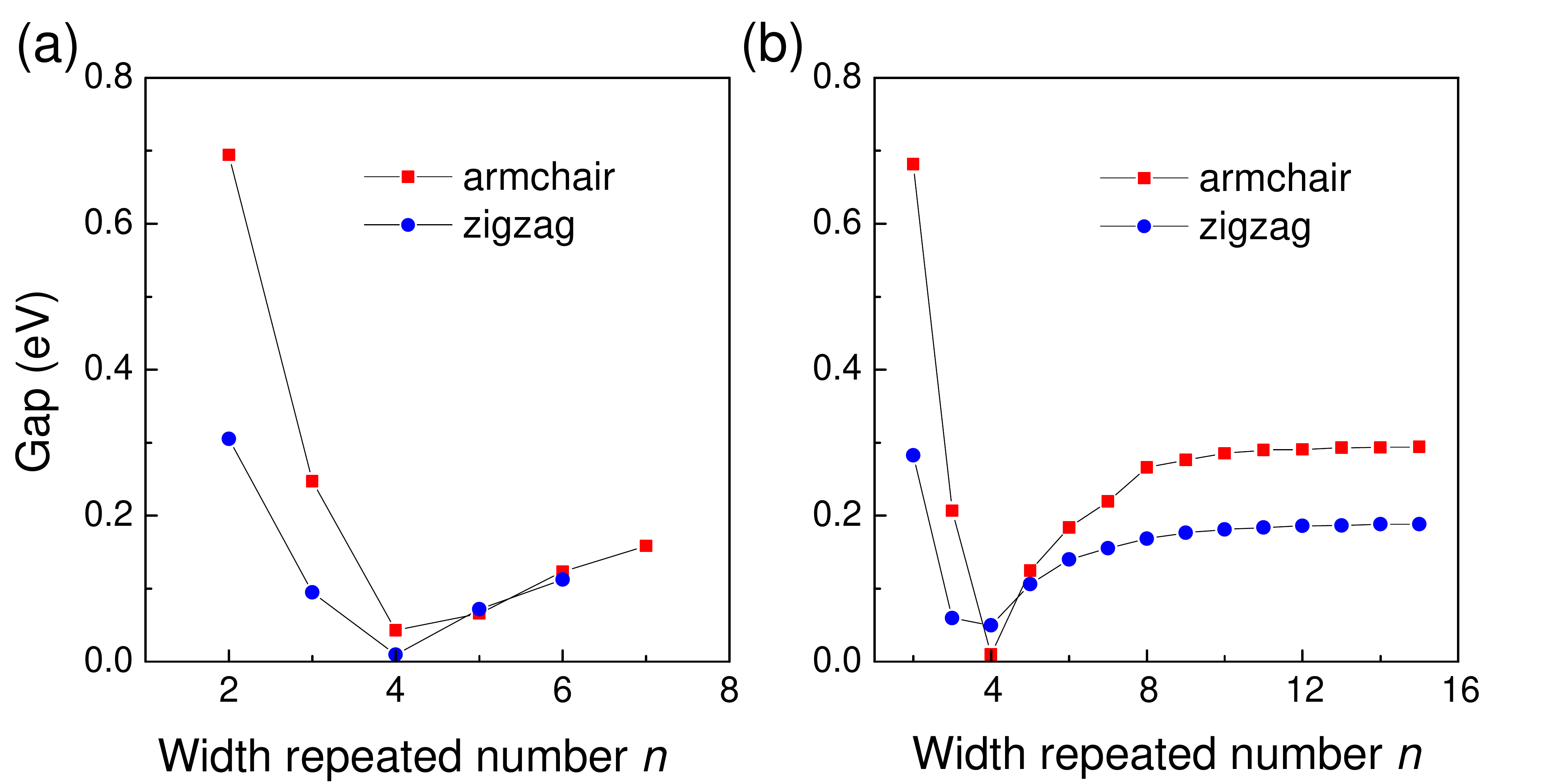}
\caption{Band gap as a function of GDYNR width, calculated by (a) VASP and (b) OpenMX, respectively.}
\label{fig:gap}
\end{figure}

For armchair GDYNR, all of them are semiconductors with a direct energy gap at $\Gamma$ point, as shown in Figs.~\ref{fig:band} (a-c) for $n=2$, $4$ and $7$. It is interesting that their energy gaps are dependent on the width. The GDYNR with $n=2$ has the largest energy gap of 0.69 eV, which is even larger than the 0.51 eV of the two-dimensional infinite graphdiyne. The $n=3$ GDYNR has an energy gap of 0.25 eV, which is quite smaller than that of infinite graphdiyne.  The $n=4$ armchair GDYNR has the smallest energy gap, 0.04 eV, which is nearly a semimetal. For $n>4$, the energy gap increases as the width increases. Fig.\ref{fig:band} (a-c) shows the electronic energy bands of armchair GDYNR-$n$ ($n=2,4,7$). For zigzag GDYNR, the energy gap for the $n=2, 3, 4, 5, 6$  nanoribbon is 0.30 eV, 0.10 eV, 0.01 eV, 0.07 eV and 0.11 eV, respectively.  A clear tendency shows that there is a minimum gap at $n=4$ for  both kinds of GDYNR, as shown in Fig.\ref{fig:gap}. The tendency was confirmed by two different DFT softwares, VASP  and OpenMX.  The energy gap 0.01 eV of $n=4$ zigzag GDYNR by VASP is the smallest, indicating that the $n=4$ zigzag nanoribbon could be metallic in room temperature.

The curve of energy gap as a function of GDYNR width is very interesting. It decreases firstly and increases again as the width increases, and reaches a minimum value at $n=4$ for both armchair and zigzag nanoribbons. To further understand the intriguing property, we construct a tight-binding (TB) model by only considering $\pi$ electrons for each atom because the PDOS analysis shows that $p_z$ orbital is the most important state around Fermi energy.  Thus, $\pi$ elctrons and nearest neighbor hoppings can be sufficient to describe the effective low-energy Hamiltonian. The TB Hamiltonian reads
\begin{equation}   
H=\sum _{ <i,j> } t _{ i,j }{ \hat { c }  }_{ i }^{ \dagger  }{ \hat { c }  }_{ j }+ \sum _{ i,j } h_{i,j}{ \hat { c }  }_{ i }^{ \dagger  }{ \hat { c }  }_{ j }
\end{equation}
where the operators ${ \hat { c }  }_{ i }^{ \dagger  } $ and ${ \hat { c }  }_{ i } $ are the creation and annihilation operators of $\pi$ electrons at site $i,<i,j>$ denotes the sum over nearest neighbors, and $t_{i,j}$ stands for the hopping amplitude between sites $i$ and $j$. The first term describes the hopping interaction within the two edges and the second term stands for the coupling between them.

\begin{figure}[tbp]
\includegraphics[width=1.0\linewidth,clip]{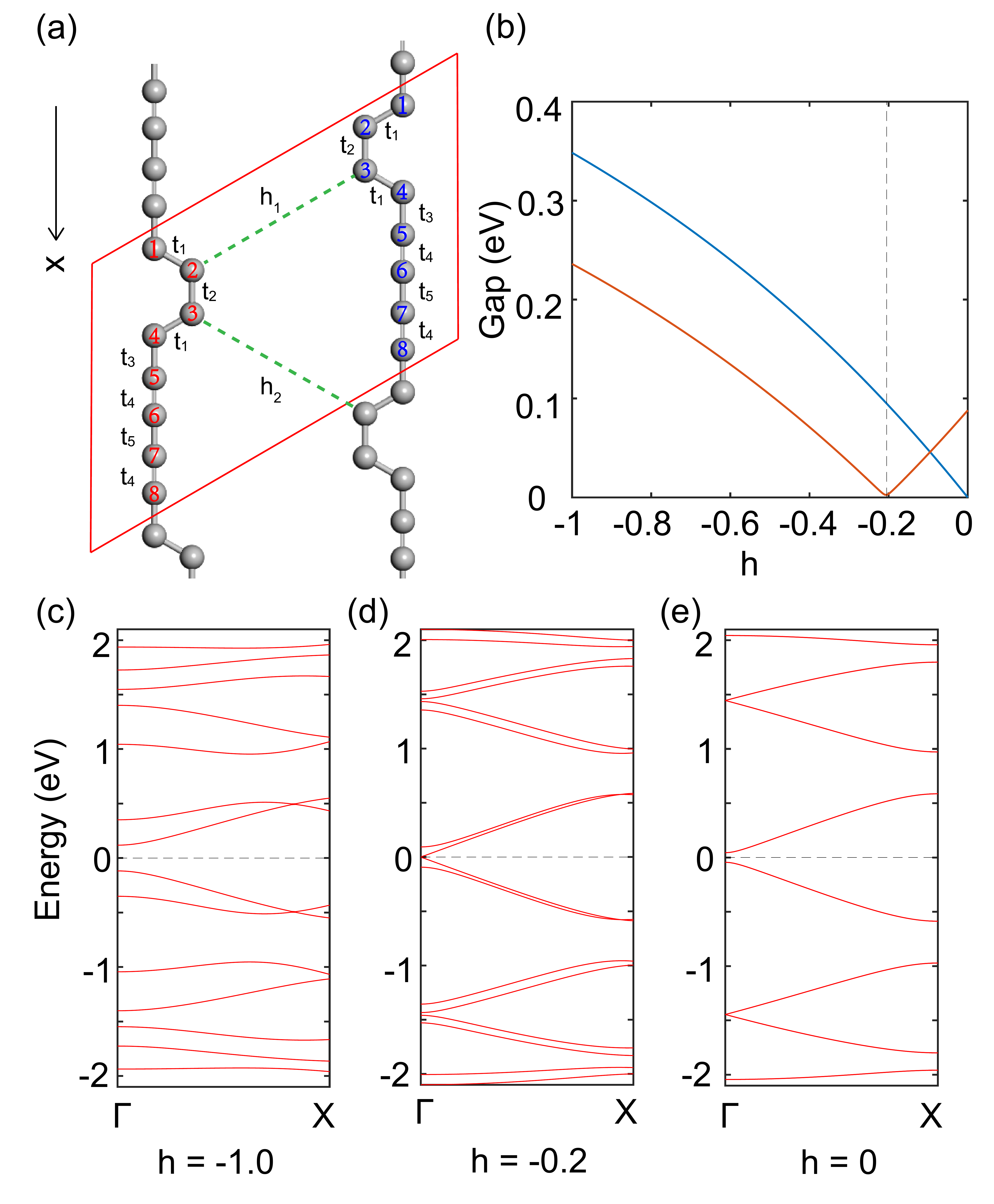}
\caption{(a) Schematic view of armchair GDYNR for construction of tight-binding model. The numbers indicates sites on edges, and $t_i$ indicates nearest neighbor hoppings of the corresponding  bond.  (b) The curve of energy gaps as a function of edge interaction hopping  $h$ with onsite energies taking the same value (blue line) and different values (red line). (c)-(e) Typical band structures by the TB model (correspond to red line in Fig. (b)) for  $h=-1.0, -0.2$ and $0$, respectively.}
\label{fig:tb}
\end{figure}

  Taking armchair GDYNR as an example, there are two edges for every nanoribbon, and each edge contains 8 carbon atoms per unit cell (see Figs.~\ref{fig:tb}(a)). The atoms in the left (right) edge are labeled as red (blue) numbers. 
Firstly, A low-energy Hamiltonian ($H_0$) of $8\times8$ size for each edge can be constructed, and there are five independent hopping parameters $t_i$ ($i=1,2,3,4,5$) [see Fig.~\ref{fig:tb} (a)]. This Hamiltonian could describle the low-energy physics of GDYNR with infinite width, because the interaction for the two edges can be neglected.  For a finite width GDYNR, the interaction between the two edges is non-ignorable.

Considering the character of the structures, the electrons can only hop along the carbon chains. Therefore, two hopping parameters $ h_{1,2}$ are enough to describe the interaction in-between. If the width is large enough, $h_{1,2}=0$. Otherwise, $h_{1,2}$ should take finite value.  After including the coupling term,  the TB Hamiltonian would be enlarged as $16\times16$ in size. For simplicity, we take the two coupling parametes as $h_1=h_2=h$. 
 Note that, if  all the onsite energies are set as the same value, the energy gap is zero for $h=0$ and increase monotonically as $|h|$ increases (see Fig.~\ref{fig:tb}). Since the atoms in a unit cell are asymmetric as well as the effect of saturated H and O atoms on the edges, the onsite energies cannot take the same value. By carefully fitting the first-principles results, the hopping parameters  can be  set as $t_1=t_2=t_3=t_4=t_5=-1$ eV, and the onsite energies take $E_1=0.3$, $E_2=-0.3$, $E_3=0.3$, $E_4=-0.3$, $E_5=-0.3$, $E_6=0.3$, $E_7=-0.3$ and $E_8=0.3$ in unit of eV, where $E_i$ is the onsite energy of atom $i$.  When the interaction parameter $h$ varies from $-1$ to $0$, the energy gap decreases firstly and then increases, and reaches a minimum value at $h=-0.2$ (see Fig~\ref{fig:tb}(b)). It gives a qualitative interpretation of the change of energy gap versus width of GDYNR. Thus, it can be concluded that the energy gap of GDYNR is dependent on the edge coupling. An isolated edge holds a finite gap, and the energy bands of the two edges are two-fold degenerate. When the interaction cannot be neglected, the two-fold degeneracy would be removed. One band is pushed up and another one is pulled down, so that the gap decreases up to a minimum. If the edge coupling were large enough, the gap would be enlarged again, quite similar to the effect of spin-orbit coupling~\cite{Sheng2014,ShengXL_JPCL2017}. Therefore,  the existence of the minimum of the energy gap is due to the competition between the interaction within the two edges and the coupling in between.
 

\begin{figure}[tbp]
\includegraphics[width=1.0\linewidth,clip]{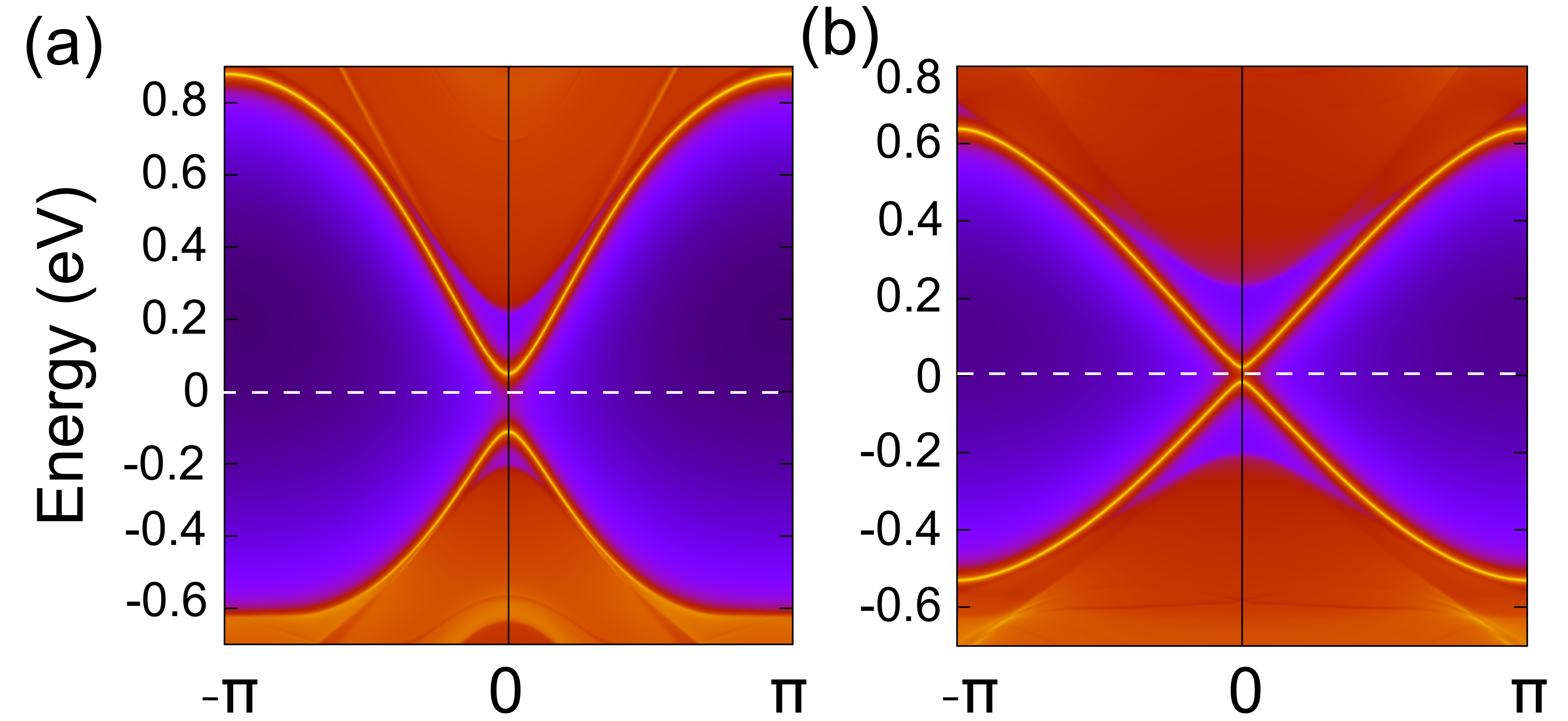}
\caption{Projected spectrum on (a) armchair and (b) zigzag edges for a semi-infinite system. Topological unprotected edge state exists in both cases, because there is a gap rather than a Dirac cone in surface states.}
\label{fig:surf}
\end{figure}

\begin{figure}[tbp]
\includegraphics[width=1.0\linewidth,clip]{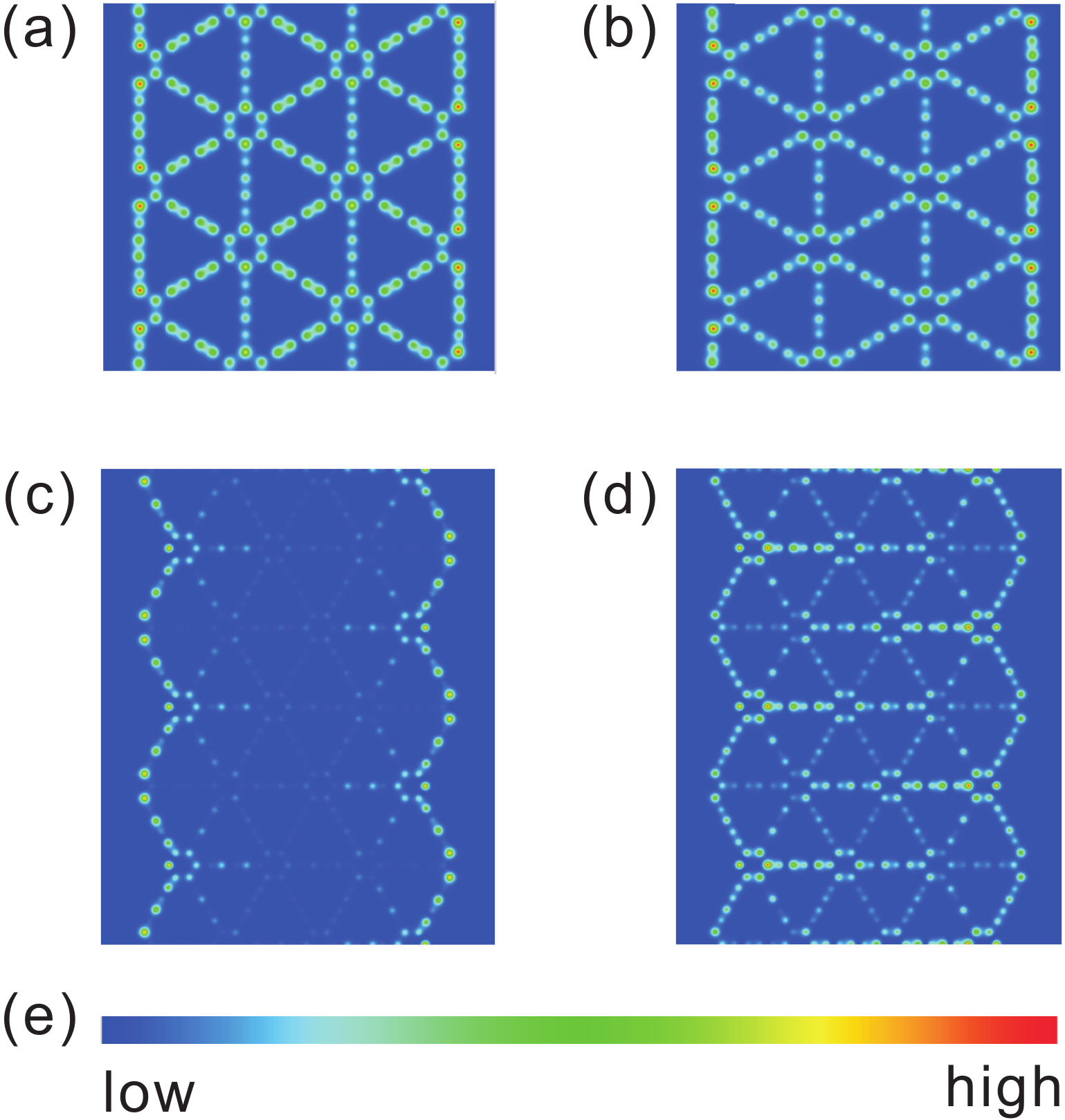}
\caption{(a) Charge density of valence band of armchair GDYNR-4 at $\Gamma$ point. (b) Hole density of conducjtion band of armchair GDYNR-4 at $\Gamma$ point. (c) Charge density of valence band of zigzag GDYNR-4 at $\Gamma$ point. (d) Hole density  of conduction band of zigzag GDYNR-4 at $\Gamma$ point.}
\label{fig:charge}
\end{figure}

To further understand the electronic properties of edge states, we construct the maximally localized Wannier functions (MLWF)~\cite{Marzari1997,Souza2001,Wannier90T} for C $s$ and $p$ orbitals, which reproduce the low energy bands accurately.  We use an iterative method~\cite{0305-4608-15-4-009,Wu_tool} to obtain the surface Green's function for the semi-infinite system. The imaginary part of the surface Green's function is the local density of states (LDOS), from which the dispersion of the surface states can be obtained. Fig.~\ref{fig:surf} shows the band structures of the semi-infinite system for both armchair and zigzag edges, from which we can find the clear edge states in the bulk gap. However, such edge states are topological unprotected, since the $\mathbb{Z}_2$ invariant of graphdiyne is zero. So that, the edge states can be moved up and down by changing the boundary conditions and result into the instability of the edge gap.  The edges states here are mainly contributed by $\pi$-electrons of carbon atoms on the edges. The analysis are confirmed by DFT calculations of the  charge (hole) density of valence (conduction) band at $\Gamma$ point.  For armchair GDYNRs, the charge density is not fully localized on the edge of the ribbons, but the charge density around the edges is larger than that of middle regions, as shown in Figs.~\ref{fig:charge} (a) and (b).  For zigzag GDYNRs,  the charge and hole density mainly spread on the two edges, especially for the charge density of valence band at $\Gamma$ point, as shown in Fig.~\ref{fig:charge} (c) and (d).  Therefore, the edge states here are different from that of topological materials~\cite{Sheng2014,ShengXL_JPCL2017,Zhang2016kk,Sheng2017,PhysRevB.94.155112,arXiv170501424}, although the edge states of TI are also localized on edges. 

\begin{figure}[tbp]
\includegraphics[width=1.0\linewidth,clip]{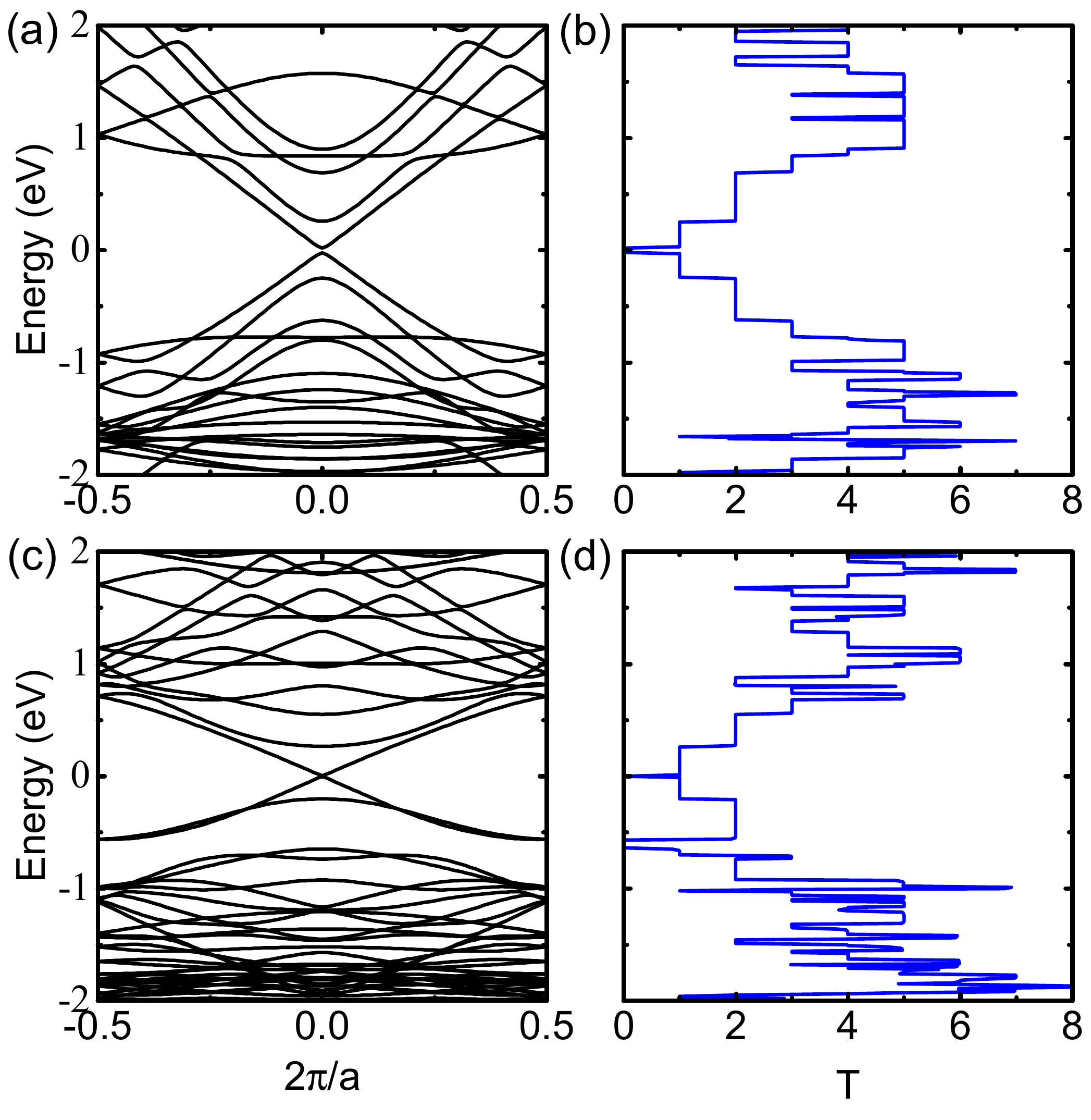}
\caption{ Band structure of (a) armchair GDYNR-4 and (c) zigzag GDYNR-4, and corresponding transmission spectrum in (b) and (d).}
\label{fig:trans}
\end{figure}

To investigate the nanoribbon conductance, we calculatated the transmission spectrum for the two GDYNR-4, armchair and zigzag, as shown in Fig.~\ref{fig:trans}. Based on a tight-binding model. we computed the real-space retarded Green's function~\cite{Datta1995}, \mbox{$\hat{G}^r(E;i,j)=[E-\hat{H}-\hat{\Sigma}^r_L -\hat{\Sigma}^r_R]^{-1}$},  the central region of graphdiyne nanoribbons attached to the left (L) and right (R) semi-infinite leads with the same width. The self-energies $\hat{\Sigma}^r_{L,R}$ are generated by leads determining escape rates of electrons from the central region into the macroscopic reservoirs kept at electrochemical potential $\mu_{L,R}$. The central region is perfectly clean in our consideration, and thus the transmission spectrum is perfectly quantized. It is clear that the conductance is zero around Fermi energy, becaused of the small gap in the nanoribbons. Away from the band gap, the transmission conductance becomes an integer, in good agreement with the band structure. Please note that althouth the edge states contribute conductance for a clean system, it is not guaranteed for a defect system because it is not topological protected~\cite{Sheng2017,2017arXiv170609361M}. 

\section{Conclusion}

In summary, we calculated the electronic structure of a novel kind of  GDYNR with open hexagonal rings on the edges. To avoid the effects of dangling bonds, hydrogen or oxygen atoms were absorbed on the edges. There are two kinds of GDYNR from the view of edge structures, armchair and zigzag. The band gap can be tuned by the width of GDYNR. As the  width of nanoribbons increases, the energy gaps decrease firstly and then increases, and  reaches a minimum gap for both of the armchair and zigzag GDYNR. A low-energy TB model for GDYNR has been constructed and concluds that the existence of the minimum of the energy gap is due to the competition between the interaction within the two edges and the coupling in between.
All the nanoribbons are direct     band gap semiconductor. The smallest gap is 0.01 eV, and therefore it could be metalic at room temperature. The largest gap is about 0.69 eV. This kind of nanoribbons hold edge states, although they are not topological insulators. The tunable band gap of GDYNR with open hexagonal rings on the edges may have a great potentail of application in nanoelectronics.

\begin{acknowledgements}
The authors thank S. A. Yang, W. Li and Z.Y. Chen for valuable discussions. 
This work is supported by the NSF of China (No. 11504013).
\end{acknowledgements}


\bibliography{draft}


\end{document}